\documentclass{aa}

\usepackage{graphicx}
\usepackage{amsmath,amssymb}
\usepackage{subfigure}

\begin{document}

\def\feii{[Fe~{\sc ii}]} \def\oi{[O~{\sc i}]} \def\sii{[S~{\sc ii}]}
\title{Near-IR \feii~emission diagnostics applied to cold disk winds
  in young stars} \subtitle{}

\author{N. Pesenti\inst{1}, C. Dougados\inst{1}, S. Cabrit\inst{2},
  D. O'Brien\inst{3}, P. Garcia\inst{3} \and J. Ferreira\inst{1} }

\offprints{N. Pesenti}

\institute{Laboratoire d'Astrophysique UMR 5571, Observatoire de
  Grenoble, BP 53, 38041 Grenoble Cedex-9, France\\ email:
  Nicolas.Pesenti@obs.ujf-grenoble.fr \and LERMA, Observatoire de
  Paris, UMR 8112 du CNRS, 61 avenue de l'Observatoire, F-75014 Paris
  \and Centro de Astrof\'\i sica da Universidade do Porto, Rua das
  Estrelas, 4150-762 Porto, Portugal }

\date{Received 9 April 2003 / Accepted 12 June 2003}

\abstract{ We investigate the emissivity properties of the main
  near-IR transitions of the Fe$^{+}$ ion in the conditions prevailing
  in the inner regions of jets from young stars, based on a simplified
  16-level atom model. We present new diagnostic diagrams involving
  prominent near-IR line ratios that allow us to constrain the electronic
  density, temperature, and Fe gas phase abundance ratio,
  independently of the heating process. Comparison with recent near-IR
  observations of a sample of HH~objects indicates gas phase
  Fe abundances ranging from 15--50~\% up to 100~\% of the solar value
  (depending on the assumed temperature and on the HH~object) in
  agreement with the moderate depletions previously derived from
  optical line ratios or shock models. Hence, it appears that
  Fe-bearing dust is efficiently destroyed in stellar jets.  We then
  use our Fe$^+$ emissivity model to predict near-IR \feii\ emission
  maps for self-similar, cold MHD disk wind models.  We show that
  near-IR \feii\ lines are stronger than \sii~$\lambda$6731 and
  \oi~$\lambda$6300 in the cool regions ($T \le 7000$~K) near the wind
  base, and that observations in \feii\ with AMBER on the VLTI could
  severely constrain the MHD solution and the inner launch radius of
  the jet. We also compare theoretical predictions with recent
  observations in the near-IR \feii\ lines of the L1551-IRS5 and
  DG~Tau jets. The cold disk wind model reproduces quite well
  the two velocity components observed at $-100$ and
  $-300$~km\,s$^{-1}$, although the high velocity component appears
  overestimated by a factor of 1.5 in the DG~Tau jet.  However, the
  model predicts too little emission at intermediate-velocity and
  insufficient densities. Similar problems were encountered in
  previous model comparisons in the optical range with jets from
  T~Tauri stars. Denser disk winds with stronger heating at the jet
  base, which have been invoked for optical jets, also appear needed
  in younger, embedded Class I jet sources.   } \authorrunning{N. Pesenti et al.}  \titlerunning{Near-IR
  \feii~emission...}  \maketitle
%

\section{Introduction}
\label{intro}

Constraining the physical mechanism by which mass is ejected from
young accreting stars and collimated into jets remains a fundamental
open problem in star formation theory. The fact that a correlation
between accretion and ejection signatures is seen from the embedded
Class~0 and Class~I phases (ages a few 10$^4$--10$^5$~yr) down to the
optically revealed T~Tauri phase (at ages a few 10$^6$~yr), with a
similar ejection to accretion ratio, suggests that the same mass-loss
mechanism may be at play throughout the different stages of star
formation (see e.g. Cabrit \cite{cabrit2002}). Studies of the inner
jet structure at each phase of evolution are required to confirm the
validity of this scenario. Attempts to compare observations and MHD
wind predictions have first concentrated on the small scale jets
associated with optically revealed T Tauri stars (Cabrit et
al. \cite{cabrit1999}; Garcia et
al. \cite{garcia2001a},\cite{garcia2001b}; Lavalley-Fouquet et
al. \cite{lavalley-fouquet}; Bacciotti et al. \cite{bacciotti2000},
Shang et al. \cite{shang1998},\cite{shang2002}). These studies are
conducted in the strong optical forbidden lines of \oi~$\lambda$6300,
\sii~$\lambda$6731 and [N~{\sc ii}]~$\lambda$6584. Thanks to recent
advances in near-IR instrumentation, similar work can now be extended
to the inner regions of jets associated with the younger more
extincted Class I sources.

  The imaging study of embedded jet sources with NICMOS/HST made by
Reipurth et al. (\cite{reipurth}) clearly demonstrated the importance
of \feii\ near-IR lines as a tracer of the innermost jet regions, less
affected by extinction than their optical counterparts. Furthermore,
Nisini et al. (\cite{nisini}) recently performed a near-IR
spectroscopic survey (from 0.95 to 2.5~$\mu$m) of a sample of HH~jets,
providing a first database to investigate the excitation conditions of
the Fe$^{+}$ ion in young stellar jets. Detailed modeling is now
needed to see how such observations can be used to constrain the jet
driving mechanism in embedded sources.

We investigate in this article the emissivity properties of the main
near-IR lines of the Fe$^{+}$ ion, based on a simplified 16-level
model, in the conditions prevailing in the inner regions of jets from
young stars. Extending the work of Nisini et al. (\cite{nisini}) and
Beck-Winchatz et al. (\cite{beck1994}), we first derive in Sect. 2
diagnostic diagrams that allow to constrain, independently of the
heating process and ejection model, the electronic
temperature and density as well as the Fe$^+$/S$^+$ gas phase
abundance ratio --- a critical parameter to derive mass loss rates
from \feii\ line fluxes and to estimate the amount of depletion onto
dust grains. We then turn in Sect. 3 to detailed predictions
for a specific class of ejection model, the self-similar cold disk
wind solutions developed by Ferreira (\cite{ferreira1997}), for which
consistent thermal solutions have been recently computed by Garcia et
al. (\cite{garcia2001a}). Our study extends the work of Garcia et
al. (\cite{garcia2001b}), which was restricted to the optical domain.
We compare these predictions with recent observations in near-IR
\feii\ lines of young stellar jets from embedded sources, in
particular the velocity resolved spectroscopy of the inner regions of
the L1551-IRS5 (Pyo et al. \cite{pyo2002}) and the DG~Tau jets (Pyo et
al. \cite{pyo2003}).  We summarize our findings in Sect. 4.


\section{Near-IR \feii\ diagnostic diagrams}

\subsection{The 16-level Fe$^+$ model}
\label{model}

\begin{figure}
        \begin{center}
        \includegraphics[scale=0.39]{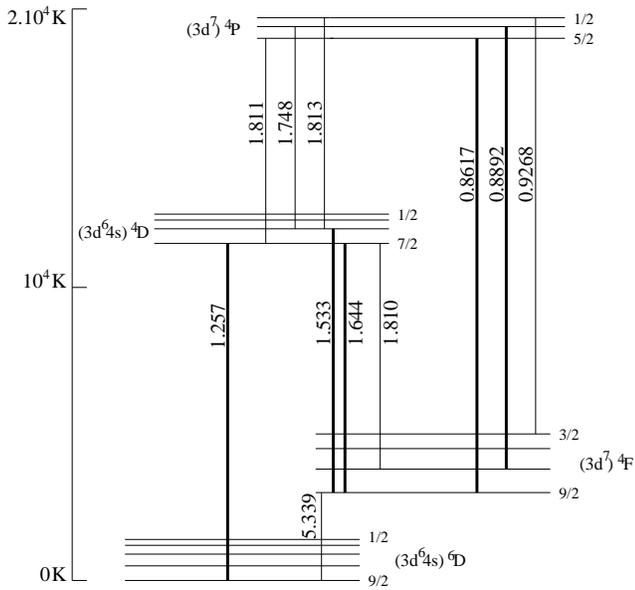}
        \caption{Energy level diagram with the prominent optical and
          near-infrared [Fe~{\sc ii}] emission lines (wavelengths in
          $\mu$m) originating from the 16 first low-lying levels. We
          also indicate on the left the range of corresponding upper
          level energies (E/$k$ expressed in K).}
        \label{level}
        \end{center}
\end{figure}

\begin{figure}
        \begin{center}
        \includegraphics[scale=0.42]{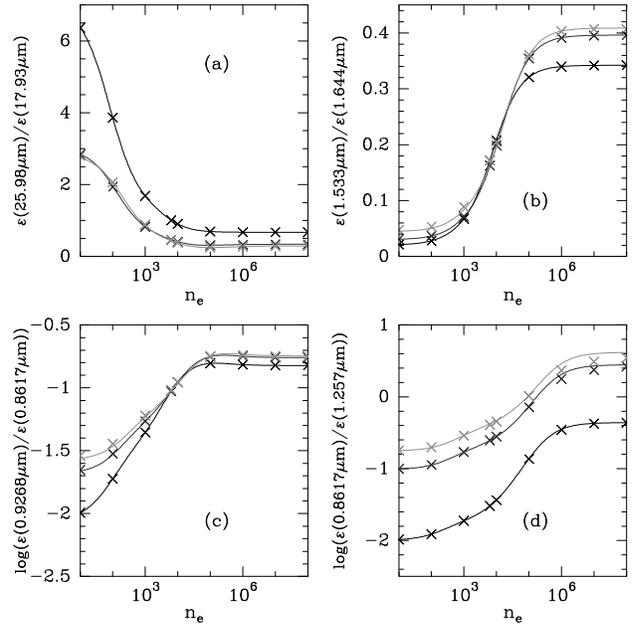}
        \caption{Theoretical line ratios as a function of n$_{\mathrm
            e}$ for different temperature values T$_{\mathrm e}$
          (3000, 10\,000, 20\,000~K). The curves are deduced from our
          16-level model and symbols from a 142-level computation
          (Pradhan, private communication). T$_{\mathrm e}$ increases
          from bottom to top in panels (b),(c) and (d), from top to
          bottom in panel (a).}
        \label{test}
        \end{center}
\end{figure}

To predict near-IR \feii\ line emissivities, we build a NLTE model
including the 16 first fine structure levels, which comprise the 4
terms (3d$^64$s)$^6$D, (3d$^7$)$^4$F, (3d$^64$s)$^4$D, (3d$^7$)$^4$P
(Fig. \ref{level}). Levels in the $^4P$ multiplet lie $\sim 1.7$~eV
above the ground state so that our simplified model will be valid for
excitation temperatures below $2 \times 10^4$~K, a condition typically
achieved in stellar jets (see below).  All radiative transitions among
these four terms are of quadrupole or magnetic dipole type. We
therefore assume optically thin conditions and compute level
populations under statistical equilibrium including electron
collisional excitation and spontaneous radiative emission
processes. We take transition probabilities from Nussbaumer \& Storey
(\cite{nussbaumer}), level energies and electron-ion collision
strengths from Zhang \& Pradhan (\cite{zhang}).  We neglect
fluorescent excitation by the UV continuum of the star. Indeed, as
pointed out by Verner et al. (\cite{verner}) in the case of the Orion
Nebula, the 16 lowest levels of the Fe$^{{\rm +}}$ ion remain
collisionally dominated in the conditions under study (optically thin
emission, n$_{\mathrm e}=1-10^8$~cm$^{-3}$, T$_{\mathrm e} \sim
10^4$~K).  Thus, the line emissivity per Fe$^{{\rm +}}$ ion only
depends on electron density (n$_{\mathrm e}$) and temperature
(T$_{\mathrm e}$).

We have tested our simplified model by comparing our results with a
more accurate \mbox{142-level} computation, taking into account the
same physical processes (Pradhan, private communication). The
comparison is shown in Fig. \ref{test} for the same four line ratios
as in Pradhan \& Zhang (\cite{pradhan}), which probe all four lowest
terms of Fe$^{{\rm +}}$. Our predicted line ratios are in very good
agreement with those obtained from the 142-level model, for n$_{\mathrm e}$
ranging from 10 to 10$^8$~cm$^{-3}$ and T$_{\mathrm e}$ ranging from 3000 to
$2\times 10^4$~K.  Differences do not exceed 1\% except for
transitions originating from the upper $^4P$ level (e.g. $0.8617\
\mu$m, $0.9268\ \mu$m) and T$_{\mathrm e}>10\,000$~K, where our model
underestimates the $0.8617\ \mu$m$/1.257\ \mu$m ratio by up to 5\,\%
for n$_{\mathrm e}$ $\leq 10^5$~cm$^{-3}$, and by $\sim 20$~\% for
n$_{\mathrm e}>10^5$~cm$^{-3}$. These differences still remain below current
observational uncertainties in line ratios. Our simplified 16-level
model is thus sufficiently accurate for diagnostic purposes.

 In the following, we concentrate on the prominent optical and
near-infrared lines. Most of the brightest transitions arise from the
$^4D$ term, namely $1.257\ \mu$m, $1.644\ \mu$m and $1.533\ \mu$m
(Fig. \ref{level}). The decay $^4F_{9/2}-^4P_{5/2}$ ($0.8617\ \mu$m)
or $^6D_{9/2}-^4F_{9/2}$ ($5.339\ \mu$m) lines also have significant
emissivities ($\geq$ one tenth of the $1.257\ \mu$m line). The latter
is inaccessible from ground-based telescopes but may be observable
from space, with JWST for example.

\subsection{Derivation of (n$_{\mathrm e}$, T$_{\mathrm e}$)}
\label{diag_diag}

\begin{figure}
        \begin{center}
        \includegraphics[scale=0.4]{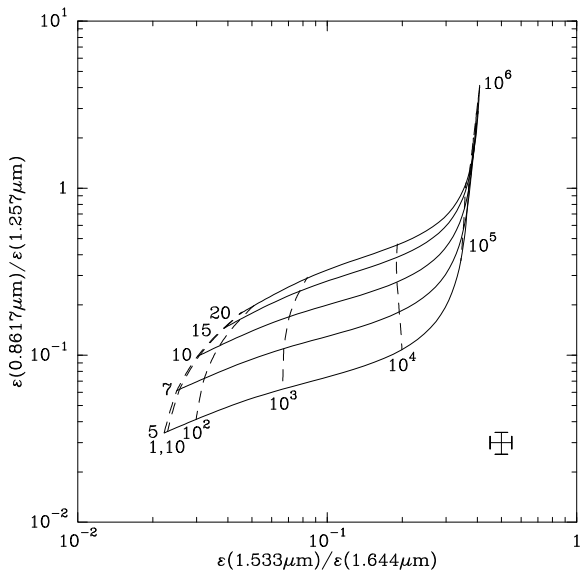}
        \caption{Diagnostic diagram of (n$_{\mathrm e}$,T$_{\mathrm
            e}$): Temperature sensitive ratio
          0.8617~$\mu$m/1.257~$\mu$m versus electronic density
          sensitive ratio 1.533~$\mu$m/1.644~$\mu$m. \textbf{Dashed
            lines}: n$_{\mathrm e}$ varying from $1$ to
          10$^6$~cm$^{-3}$ in factors of 10. \textbf{Solid lines}:
          T$_{\mathrm e}$ varying from 5000 to 20\,000\,K (T$_{\mathrm
            e}$ expressed in factors of 10$^3$~K). Error bars
          correspond to typical uncertainties of $\pm$\,10~\% and
          $\pm$\,15~\% in 1.533~$\mu$m/1.644~$\mu$m and
          0.8617~$\mu$m/1.257~$\mu$m respectively.}
        \label{isorap}
        \end{center}
\end{figure}

\begin{table}
  \begin{tabular}{l l c c}
    \hline \noalign{\smallskip} \vspace{0.2cm} Level & Transition &
    \multicolumn{2}{c}{n$_{cr}$~(cm$^{-3}$)} \\ & & 5000\,K &
    10\,000\,K \\ \hline \noalign{\smallskip} \feii
    $\,\phantom{}^4D_{9/2}$ & $1.644,1.257\ \mu$m & $3.5\, 10^4$ &
    $5.6\, 10^4$ \\ \feii $\,\phantom{}^4D_{7/2}$ & $1.533\ \mu$m &
    $2.9\, 10^4$ & $4.6\, 10^4$ \\ \feii $\,\phantom{}^4P_{5/2}$ &
    $0.8617\ \mu$m & $2.6\, 10^5$ & $3.5\, 10^4$ \\ \oi
    $\,\phantom{}^1D_{2}$ & $0.6300\ \mu$m & $4.6\, 10^6$ & $1.8\,
    10^6$ \\ \sii $\,\phantom{}^2D_{3/2}$ & $0.6731\ \mu$m & $2.7\,
    10^3$ & $3.5\, 10^3$ \\
\hline
  \end{tabular}
  \caption{Critical densities (n$_{cr}$) of upper levels
for the prominent atomic transitions, calculated using Eq. (1) for
T$_{\mathrm e}$= 5000~K and 10\,000~K}
  \label{ncr}
\end{table}

Critical densities for the most intense \feii\ lines, and for the main
forbidden lines of \sii\ and \oi\ observed in jets, can be calculated
with the generalized formula of Osterbrock (\cite{osterbrock}):
\begin{equation}
n_{cr} (i) = \Sigma_{j<i}~A_{ij}/\Sigma_{j \ne i}~q_{ij},
\end{equation} 
where A$_{ij}$ denotes Einstein spontaneous emission coefficients and
q$_{ij}$ denotes collisional rate coefficients.  The results are given
in Table \ref{ncr}, and show that the \feii\ lines investigated here
have critical densities of a few $10^4$~cm$^{-3}$ (slightly higher
than the red \sii~$\lambda\lambda$6716,6731 lines \footnote{The
critical density for the \sii~$\lambda$6731 line in Table \ref{ncr} is
markedly lower than the value $\simeq 10^4$~cm$^{-3}$ often quoted in
the literature, derived from a two-level atom approximation.}). They
will therefore probe emitting regions with electronic densities of the
same order. Line ratios involving transitions with similar excitation
energies and similar ground states can be used to derive n$_{\mathrm
e}$. For n$_{\mathrm e}$ in the range 10$^2$ to 10$^5$~cm$^{-3}$, the
best diagnostic is the \feii\ $1.644\ \mu$m/$1.533\ \mu$m
ratio. Indeed, this ratio involves two of the brightest near-IR \feii\
lines and is little dependent on temperature and reddening
(Fig. \ref{test}b).  Note that the brightest two transitions at 1.644
and 1.257~$\mu$m originate from the same upper level. Their ratio
($\epsilon(1.257\ \mu$m)/$\epsilon(1.644\ \mu$m)=1.36) is therefore
insensitive to the excitation conditions and can be used to derive the
extinction A$_{\rm V}$ along the line of sight.
 
Since most of the prominent near-infrared transitions have similar
excitation energies, their ratio is not very sensitive to the gas
temperature. Combining them with transitions from the $^4P$ term, with
higher excitation energy, could in principle lead to gas temperature
determinations. In the near-infrared domain, the transitions at
1.811~$\mu$m and 1.813~$\mu$m would be very good candidates. However,
these lines lie in a region of poor atmospheric transmission and in
addition are blended with the stronger $^4F-^4D$ $1.810\ \mu$m
line. The line at $1.748\ \mu$m is weaker and may be, at the present
available spectral resolutions, strongly contaminated by the H$_2$ 1-0
S(7) transition (cf.  Nisini et al. \cite{nisini}).  A few lines lying
in the red part of the optical spectrum can be used, the most
prominent being at 0.8617~$\mu$m and 0.8892~$\mu$m. In
Fig. \ref{isorap}, we propose a diagnostic diagram, combining the
density sensitive \feii\ $1.644\ \mu$m/$1.533\ \mu$m ratio with the
temperature sensitive \feii\ $0.8617\ \mu$m/$1.257\ \mu$m ratio, which
provides a direct measure of (n$_{\mathrm e}$,T$_{\mathrm e}$). A similar diagram,
including the red transition at $^4F_{3/2}-^4P_{3/2}$ $0.9471\ \mu$m,
is proposed in Nisini et al. (\cite{nisini}). With a flux
ratio uncertainty of $\pm$\,10~\% for the \feii~$1.644\ \mu$m/$1.533\
\mu$m ratio, the diagnostic diagram presented in Fig. \ref{isorap}
would allow determination of n$_{\mathrm e}$ ranging from $10^2$ to
$10^5$~cm$^{-3}$ with less than a factor 4 uncertainty (total
error). To achieve an uncertainty of $\pm$\,15~\% on T$_{\mathrm e}$ at
T$_{\mathrm e} \lesssim 10\,000$~K, an estimated error of $\pm$\,15~\% in
the \feii~$0.8617\ \mu$m/$1.257\ \mu$m ratio is required.

\subsection{Diagnostic of gas phase Fe abundance}
\label{Fedep}

\begin{figure}
        \begin{center}
        \includegraphics[scale=0.4,angle=-90]{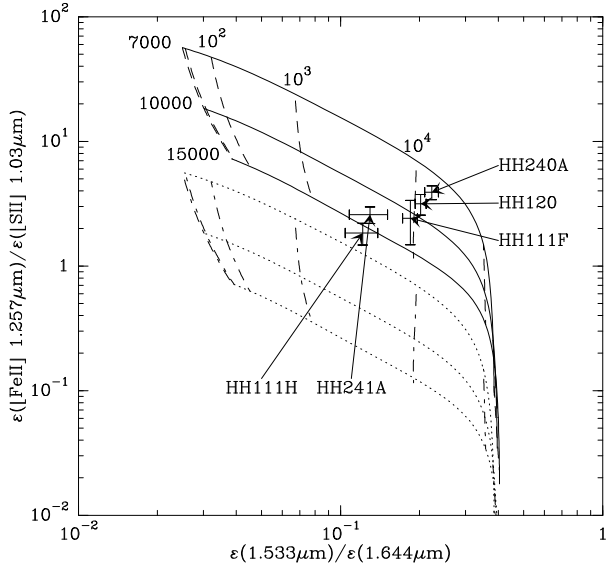}
        \caption{Diagnostic diagram of the Fe$^{\rm +}$/S$^{\rm +}$ gas phase
        abundance. We plot the density sensitive (\feii\ $1.533\
        \mu$m/$1.664\ \mu$m) ratio versus the depletion sensitive
        \feii~$1.257\ \mu$m/([S~{\sc
        ii}]~(1.029+1.032+1.034+1.037)~$\mu$m ratio. Predicted line
        ratios are plotted as function of T$_{\mathrm e}$ (solid lines) and
        n$_{\mathrm e}$ (dashed lines) for Fe$^{\rm +}$/S$^{\rm +}$ = 1.7
        (corresponding to solar abundances, Savage \& Sembach
        \cite{savage}).  Dotted and dash-dotted lines correspond to
        Fe$^{\rm +}$/S$^{\rm +}$ = 0.17 (1/10 of solar
        value). Observed line ratios for the sample of HH~objects
        studied by Nisini et al. (\cite{nisini}), corrected for individual
        reddenings, are represented with their error bars (including
        A$_{\rm V}$ uncertainties).}
      \label{depletion}
    \end{center}
\end{figure}

In the interstellar medium, condensation processes onto dust grains
deplete the gas phase abundance of heavy elements such as Fe by a
factor $> 10$ (Savage \& Sembach \cite{savage}). In jets from young
stellar objects, sputtering processes in sufficiently fast shocks (see
e.g. Jones et al. \cite{jones}) could destroy a significant fraction
of the dust grains, replenishing the gas phase with heavy
elements. Strong heating at the jet base could also evaporate dust
grains in the ejected gas. The gas phase abundance of Fe in HH~flows
is therefore an important parameter giving interesting indications on
the shock and thermal history of ejected matter. In addition,
estimates of the Fe$^{\rm +}$ gas phase abundance in HH~flows are
critically required to derive mass-loss rates from \feii\ line fluxes.

A first method to assess the iron gas phase depletion consists in
comparing the Fe abundance to that of sulfur, an element not
significantly depleted onto dust grains, with a known gas-phase
abundance of $1.86 \times 10^{-5}$ (Savage \& Sembach \cite{savage}).
In young stellar jets, emission of these elements is dominated by
\feii\ and \sii\ lines. However, the Fe$^{\rm +}$/S$^{\rm +}$ ratio derived
from these lines is probably very similar to the total Fe/S ratio. If
the lines are formed in cooling flows behind shock waves, one expects
Fe$^{\rm +}$/Fe $\le$ S$^{\rm +}$/S $\le$ 2Fe$^{\rm +}$/Fe
(Beck-Winchatz et al. \cite{beck1996}), so that Fe$^{\rm +}$/S$^{\rm
+}$ $\le$ Fe/S $\le$ 2Fe$^{\rm +}$/S$^{\rm +}$.  If jet heating is
instead dominated by ambipolar diffusion and UV radiation from the
accretion shock, S and Fe should be both predominantly in their singly
ionized form (see Garcia et al. \cite{garcia2001a}), so that Fe/S $=$
Fe$^{\rm +}$/S$^{\rm +}$.  In either case, the Fe$^{\rm +}$/S$^{\rm
+}$ ratio provides a valuable lower limit to the actual Fe/S ratio,
within a factor of 2 of the true value.

This method was first applied to HH~objects by Beck-Winchatz et
al. (\cite{beck1994},\cite{beck1996}), using a large sample of optical
\feii\ lines. The Fe$^{\rm +}$/S$^{\rm +}$ ratio is derived from the
relative intensities of \feii\ and \sii\ optical lines by means of a
statistical equilibrium model, similar to that outlined in Sect. 2.1,
given appropriate values of (n$_{\mathrm e}$,T$_{\mathrm e}$). The assumption of a uniform
emission region is partly justified by the fact that the lines are
excited under rather similar conditions (see Sect. 2.2). Inferred gas
phase Fe$^{\rm +}$/S$^{\rm +}$ ratios range from 1 to 4 when the
\sii~$\lambda\lambda$6716,6731 lines are used for reference.  When
compared with the most recent estimate of the solar Fe/S ratio
$\simeq$~1.7, consistent with the local ISM (Savage \& Sembach
\cite{savage}; Sofia \& Meyer \cite{sofia}), these values indicate no
Fe gas phase depletion in most HH~flows. Surprisingly, no
correlation with the degree of excitation of the flow is
observed. This analysis was however limited to the case of optically
revealed parts of the jets, i.e. usually far from the driving source.

Recently, Nisini et al. (\cite{nisini}) proposed two other methods to
measure the Fe gas phase abundance in young stellar jets, relying only
on near-infrared lines. The first one involves the \feii~$1.257\
\mu$m/Pa$\beta$ ratio, assuming that Pa$\beta$ is emitted under case B
recombination. A drawback of this method is that, in addition to
n$_{\mathrm e}$ and T$_{\mathrm e}$, it requires knowledge of the
hydrogen ionization fraction, x$_e$, which is not available for all
sources. Furthermore, the assumption that Pa$\beta$ is formed in the
same region as \feii, and that it is dominated by recombination, is
generally not valid in shocks (cf. Bacciotti \& Eisl\"offel
\cite{bacciotti1999}). A second method proposed by these authors is to
compare the observed near-IR line ratios \feii~$1.257\ \mu$m/[C {\sc
i}]~$0.98\ \mu$m and [Fe {\sc ii}]~$1.257\ \mu$m/Pa$\beta$ with
predictions from models of fast dissociative shocks. In the sample of
objects studied by Nisini et al.  (\cite{nisini}), the inferred gas
phase Fe/H ranges from 30~\% to 70~\% of the solar photospheric value
((Fe/H)$_{\odot}$ = $3.2 \times 10^{-5}$, Savage \& Sembach
\cite{savage}). These results depend on the assumption that the carbon
abundance in the shock remains constant at $2.3 \times 10^{-4}$,
i.e. 64~\% of the solar value (Hollenbach \& McKee \cite{hollenbach},
Savage \& Sembach \cite{savage}), while Fe is returned to the gas
phase. A solar gas-phase carbon abundance would increase their
estimated Fe/H by typically a factor 1.5.

We propose here a third approach to estimate the gas phase Fe
abundance of HH~flows from near-IR observations, independent of both
x$_{e}$ and shock models. The method is simply an extension to the
near-IR domain of that introduced by Beck-Winchatz et
al. (\cite{beck1994}) in the optical range. As described above, it
will yield a lower limit to the Fe/S ratio.  We propose to use the
ratio of the \feii~$1.257\ \mu$m line to the
\sii~$1.029+1.032+1.034+1.037\ \mu$m lines (hereafter referred to as
\sii~$1.03\ \mu$m). We predict comparable integrated flux in these
lines in the conditions under study and, indeed, these near-IR \sii\
lines have been detected by Nisini et al. (\cite{nisini}) in their
sample of HH objects. We have chosen the \feii~$1.257\ \mu$m line
because it is one of the brightest transitions and it minimizes
reddening uncertainties relative to \sii~$1.03\ \mu$m. To compute the
emissivity per ion for S$^{\rm +}$, we construct a model including the
5 low-lying metastable levels. Transition probabilities are taken from
Fritzsche et al. (\cite{fritzsche}), collision strengths from Cai \&
Pradhan (\cite{cai}). We also use the \feii~$1.533\ \mu$m/$1.644\
\mu$m ratio, as an indicator of n$_{\mathrm e}$. Fig.~\ref{depletion}
presents the resulting diagnostic diagram, which plots \sii~$1.03\
\mu$m/\feii~$1.257\ \mu$m versus \feii~$1.533\ \mu$m/$1.644\ \mu$m,
and yields the gas phase Fe$^{\rm +}$/S$^{\rm +}$ ratio provided
T$_{\mathrm e}$ is known.  It can be seen that near-IR \feii/\sii\
line ratios are more sensitive to temperature than the optical ratios
(see Table 2 of Beck-Winchatz et al. \cite{beck1994}), though they are
less sensitive to reddening.

We plot in Fig. \ref{depletion} the observational data from Nisini et
al. (\cite{nisini}). Temperatures are not available for these sources,
but a range of 7000 to 15\,000~K can be safely assumed, based on the
predicted temperature of the low excitation forbidden line emitting
region in planar shock models on the one hand (Hartigan et
al. \cite{hartigan1994}), and maximum derived temperatures in
optically visible HH~flows, on the other hand (B\"ohm \& Solf
\cite{bohm}).  We can derive an absolute {\it lower limit} to the
Fe$^{\rm +}$/S$^{\rm +}$ gas phase ratio, and hence to the Fe/S ratio,
by assuming T$_{\mathrm e} = 7000$~K. These limits are: $0.25 \pm 0.1$
for HH~111H, $0.39 \pm 0.1$ for HH~241A, $0.55 \pm 0.2$ for HH~111F,
$0.83 \pm 0.15$ for HH~120H, and $1.2 \pm 0.25$ for HH~240A, and range
from $\simeq$~15~\% to 70~\% of the solar value ($\mathrm{Fe/S=1.7}$).
These values are remarkably similar to the results found by Nisini et
al. (\cite{nisini}) from shock models, with the same ordering for the
5 individual sources. However, we stress that the above estimates are
lower limits. With T$_{\mathrm e}\simeq 10^4$~K, the data are
compatible with no depletion in most sources, except HH~241A and
HH~111H (50~\% depletion). In the following, we will thus make the
assumption that all iron is in the gas phase. However, we point out
that to derive iron depletion and jet mass-loss rates to better than a
factor $\simeq$ 5 from \feii\ line fluxes, T$_{\mathrm e}$ estimates
will be critically required.

\section{Predictions for cold MHD disk winds}

In this section, we use our simplified Fe$^{\rm +}$ model to predict
near-IR \feii\ emissivities for the class of self-similar MHD cold
disk wind models developed by Ferreira (\cite{ferreira1997}), for
which fully consistent thermal solutions have been recently computed
(Garcia et al. \cite{garcia2001a}). We extend here the work presented
in Garcia et al. (\cite{garcia2001b}), which was restricted to
predictions in the optical domain. We first briefly describe the model
and compare predicted emissivities in the near-IR \feii\ lines with
those of the prominent optical \oi~$\lambda$6300 and
\sii~$\lambda$6731 emission lines. We then compare these predictions
with recent observations in the near-IR domain of the stellar jets
from L1551-IRS5 and DG~Tau.

\subsection{Wind model}
The MHD accretion-ejection structures of Ferreira
(\cite{ferreira1997}) represent an improvement on the centrifugal MHD
disk winds first calculated by Blandford \& Payne
(\cite{blandford}). Both assume that (1) matter is ejected along a
large scale bipolar magnetic field threading through the accretion
disk, (2) jet enthalpy is negligible for accelerating the flow (the
wind is "cold"), and (3) the structure is steady-state, axisymmetric,
and self-similar to the disk radius.  However, the solutions of
Ferreira (\cite{ferreira1997}) also describe self-consistently the
accretion flow in the underlying resistive Keplerian disk, including
all relevant dynamical terms that govern the wind mass-loading at the
slow-point (see Ferreira \& Pelletier \cite{ferreira1995}). This
global description imposes additional constraints on the self-similar
structure, which is then specified by only three dimensionless
parameters: (1) $\xi \equiv {\rm dlog} {\dot M}_{acc}/{\rm dlog} r$,
the ejection efficiency parameter, which is related to the total
ejection/accretion ratio by ${\dot M}_{ejec}/{\dot M}_{acc} = \xi
\times ln(r_e/r_i)$, where $r_e$ and $r_i$ are the outer and inner
radii of the MHD disk-wind structure, (2) $\epsilon = h/r_0$, where
$h$ is the disk scale height at the cylindrical radius $r_0$, and (3)
$\alpha_m = \nu_m/V_Ah$, where $\nu_m$ is the required turbulent
magnetic diffusivity and $V_A$ is the Alfvén velocity on the disk
midplane.

Following Cabrit et al. (\cite{cabrit1999}), we will adopt $\epsilon
=0.1$ (as estimated in HH~30 by Burrows et al. \cite{burrows}) and
$\alpha_m =1$. Solutions that extend far from the Alfvén surface are
then found for $\xi$ between 0.005 and 0.012 (Ferreira 1997). We
concentrate here on the solution with $\xi$ = 0.01 (Model A in Cabrit
et al. \cite{cabrit1999}), which is found to best reproduce the
collimation properties of T~Tauri microjets (Cabrit et
al. \cite{cabrit1999}; Dougados et al. \cite{dougados}). As a
result of toroidal magnetic fields, the streamlines slowly recollimate
beyond z/r$_0 \simeq 1500$ (where r$_0$ is the initial radius of the
streamline in the disk), and terminate at $z/r_0\simeq$ 8000, due to
strong refocusing towards the axis.  One expects that a more gradual
recollimation would occur in reality, due to additional pressure on
the flow axis from open stellar magnetic field lines.

\subsection{Predicted {\rm \feii}\ emissivities}

The thermal and ionization structure of the MHD disk wind
solutions of Ferreira (\cite{ferreira1997}) was computed \textit{a
posteriori} by Garcia et al. (\cite{garcia2001a}) for an atomic gas of
solar composition, along 11 streamlines anchored in the disk at radii
ranging from r$_0 = 0.07$~AU, the typical disk corotation radius for a
T~Tauri star, to 1~AU, where molecules should start to form. The
dominant heating mechanism is ambipolar diffusion, while adiabatic
expansion dominates the cooling.  We will use their results for a
central stellar mass M$_{\ast}$~=~0.5~M${_\odot}$ and accretion rates
$\dot{M}_{acc}=10^{-5}-10^{-6}$~M$_{\odot}$\,yr$^{-1}$.  We compute
the emissivities in the \feii\ near-IR lines along each streamline
with the 16-level atom model described in Section 2.1, using the
n$_{\mathrm e}$, T$_{\mathrm e}$ and Fe$^{\rm +}$/Fe ratios provided by the thermal
solution. The emissivity grid is projected onto the plane of the sky
to construct emission maps and long slit spectra. Maps are then
convolved by a two-dimensional Gaussian beam, representative of the
instrumental spatial and spectral resolutions.

\begin{figure}
  \begin{center}
    \includegraphics[scale=0.45]{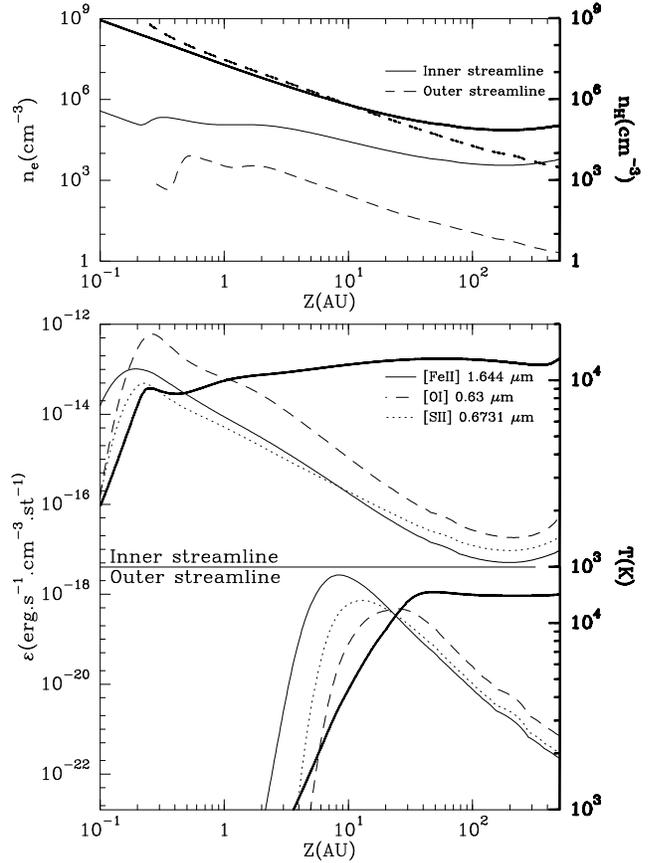}
    \caption{ {\bf Top}: Variation of n$_{\mathrm e}$ and n$_H$ (bold curve) along
      the inner (r$_{0}=0.07$~AU, full lines) and outer (r$_{0}=
      1$~AU, dashed lines) streamlines of the fiducial cold disk wind
      solution ($\xi=0.01$,
      $\dot{M}_{acc}=10^{-6}$~M$_{\odot}$\,yr$^{-1}$,
      M$_{\ast}$~=~0.5~M$_{\odot}$) as a function of the altitude Z
      above the disk.  {\bf Middle and Bottom}: Variation of line
      emissivities along the inner and outer streamlines for the 3
      emission lines: \feii~$1.644\ \mu$m, \sii~$\lambda$6731 and
      \oi~$\lambda$6300. The variation of T$_{e}$ is also shown in
      bold curve.}
    \label{emline}
  \end{center}
\end{figure}

We first compare predicted near-IR \feii\ emissivities along the flow
with those predicted for the main optical emission lines of \sii\ and
\oi. We plot in Fig. \ref{emline} the variation of emissivity with
altitude along the innermost and outermost streamlines for the
\feii~$1.644\ \mu$m, \sii~$\lambda$6731 and \oi~$\lambda$6300 lines.
Both streamlines follow the same behavior in all 3 lines: emissivity
first increases due to the central rise in temperature, the plasma
then reaches a temperature plateau ($\sim 10^4$~K) and the drop in
n$_{\mathrm e}$, due to the widening of the jet, produces the subsequent
emissivity decrease. The final increase in the inner streamline at
Z~$>$~200~AU is due to compression in the recollimation zone. Since
solutions have a self-similar geometry, maximum emissivity is reached
at a lower altitude (Z~$<$~1~AU) in the inner streamline as compared
to the outer one (Z~$\sim$~10~AU).

There are however important differences in emissivity profile between
the 3 lines close to the central source.  The upper level energies of
near-IR \feii~lines (E/$k$\,$\sim 10\,000$~K) are lower than for
\sii~$\lambda 6731$ and \oi~$\lambda 6300$ lines (E/k~$\sim
20\,000$~K), so that the \feii/\sii\ and \feii/\oi~ratios increase
dramatically at the moderate electron temperatures $\le 7000$~K
prevailing at the wind base. \feii\ lines therefore appear to be a
choice tracer for the cooler, dense regions at the wind base,
typically below a few AUs. In addition, these near-IR lines would
allow to better trace extincted regions at the base of embedded young
stellar jets.

\begin{figure}
  \begin{center}
    \includegraphics[height=88mm,angle=-90]{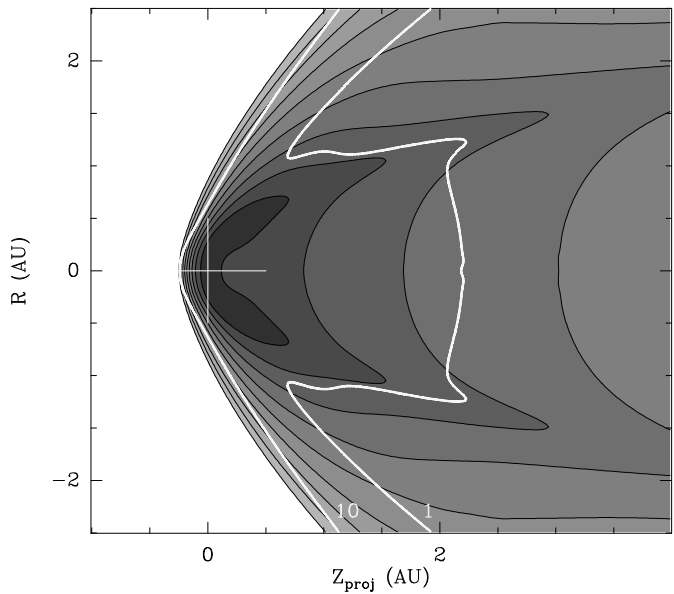}
    \caption{Predicted \feii~$1.644\ \mu$m emission map of the fiducial
     cold disk wind solution with $\xi = 0.01$,
     $\rm{\dot{M}_{acc}}$=10$^{-6}$~\mbox{M$_{\odot}$\,yr$^{-1}$} and
     inclination to the line of sight $i$ of $45^{\circ}$. The field
     of view (0.04$^{\prime\prime}$) and the Gaussian beam (1 mas,
     i.e. 0.14~AU) correspond to the AMBER/VLTI instrument
     characteristics.  Filled contours start at the maximum surface
     brightness ($2.3\,10^{-4}$~W\,m$^{-2}$\,st$^{-1}$) and decrease
     by factors of 2. White contours correspond to \feii~$1.644\
     \mu$m/\sii~$\lambda$6731 ratios of 10 and 1 (from left to
     right). The shape of the innermost streamline, assumed here
     anchored at 0.07~AU, is fully resolved and would set strong
     constraints on proposed MHD models.}
    \label{AMBER}
    \end{center}
\end{figure}

The AMBER/VLTI instrument, with a spatial resolution of $\simeq$~1~mas
(0.14~AU in Taurus) in the near-IR, will be ideally suited for such
studies of innermost jet regions.  We plot in Fig. \ref{AMBER} the
predicted AMBER/VLTI map in \feii~$1.644\ \mu$m for our fiducial cold
disk wind model, with $\dot{M}_{acc}=10^{-6}$~\mbox{M$_{\odot}$\,yr$^{-1}$}
and inclination angle to the line of sight of 45$^{\circ}$. We also
show as white contours the ratio of the \feii~$1.644\ \mu$m to the
\sii~$\lambda$6731 surface brightnesses. As expected, the \feii~line
is several times stronger than the \sii~one on scales $\leq$~2~AU
above the disk, due to the drop in T$_{e}$. At this spatial
resolution, the \feii\ map is dominated by the innermost streamline,
which is fully resolved, and shows a clear central hole. Hence, such
observations could provide direct constraints on the shape (Z(r)) of
the streamlines in the MHD jet solution and on the innermost radius at
which ejection occurs.

\subsection{Comparison with L1551-IRS5 and DG~Tau}

 In this section, we discuss constraints on the ejection mechanism
 brought by recent observations in near-IR \feii\ lines of the
 L1551-IRS5 and DG~Tau jets. We compare these observations with
 predictions from our fiducial cold disk wind model and with
 constraints derived from previous studies of T~Tauri jets in the
 optical domain.

\subsubsection{Morphology and kinematics}
\label{morphology}

 One of the best-known young stellar jets is associated with the
embedded binary source L1551-IRS5. It was first detected in the
optical by Cohen et al. (\cite{cohen}). High resolution images show
two extended ridges, identified as two separate jets with differing
line of sight velocities (Fridlund \& Liseau \cite{fridlund}; Itoh et
al. \cite{itoh}). Pyo et al. (\cite{pyo2002}) present
velocity-resolved \feii~$1.644\ \mu$m spectra towards the base of the
northern jet.  These observations provide detailed morphological and
kinematical information on the inner 500~AU of the jet, which can be
directly compared to predictions from the cold disk wind model
investigated here.

We adopt for the model an accretion rate of
$\dot{M}_{acc}=10^{-5}$~M$_{\odot}$\,yr$^{-1}$ close to the estimated
value for the L1551-IRS5 source of $6 \times
10^{-6}$~M$_{\odot}$\,yr$^{-1}$ (Momose et al.  \cite{momose}), an
inclination angle to the line of sight of $45^{\circ}$ (Pyo et al.
\cite{pyo2002}), and a central stellar mass of 0.5~M$_\odot$,
i.e. roughly half the total mass of the binary system $\simeq
1.2$~M$_\odot$ estimated from orbital motions (Rodriguez et
al. \cite{rodriguez}). Predicted emission maps and position-velocity
(PV) diagrams are convolved with Gaussian beams of widths
0.3$^{\prime\prime}$ and 60~km\,s$^{-1}$, corresponding to the
instrumental spatial and spectral resolutions of the L1551
observations.

\begin{figure}[!ht]
  \centerline{\includegraphics[height=0.99\columnwidth,angle=-90]{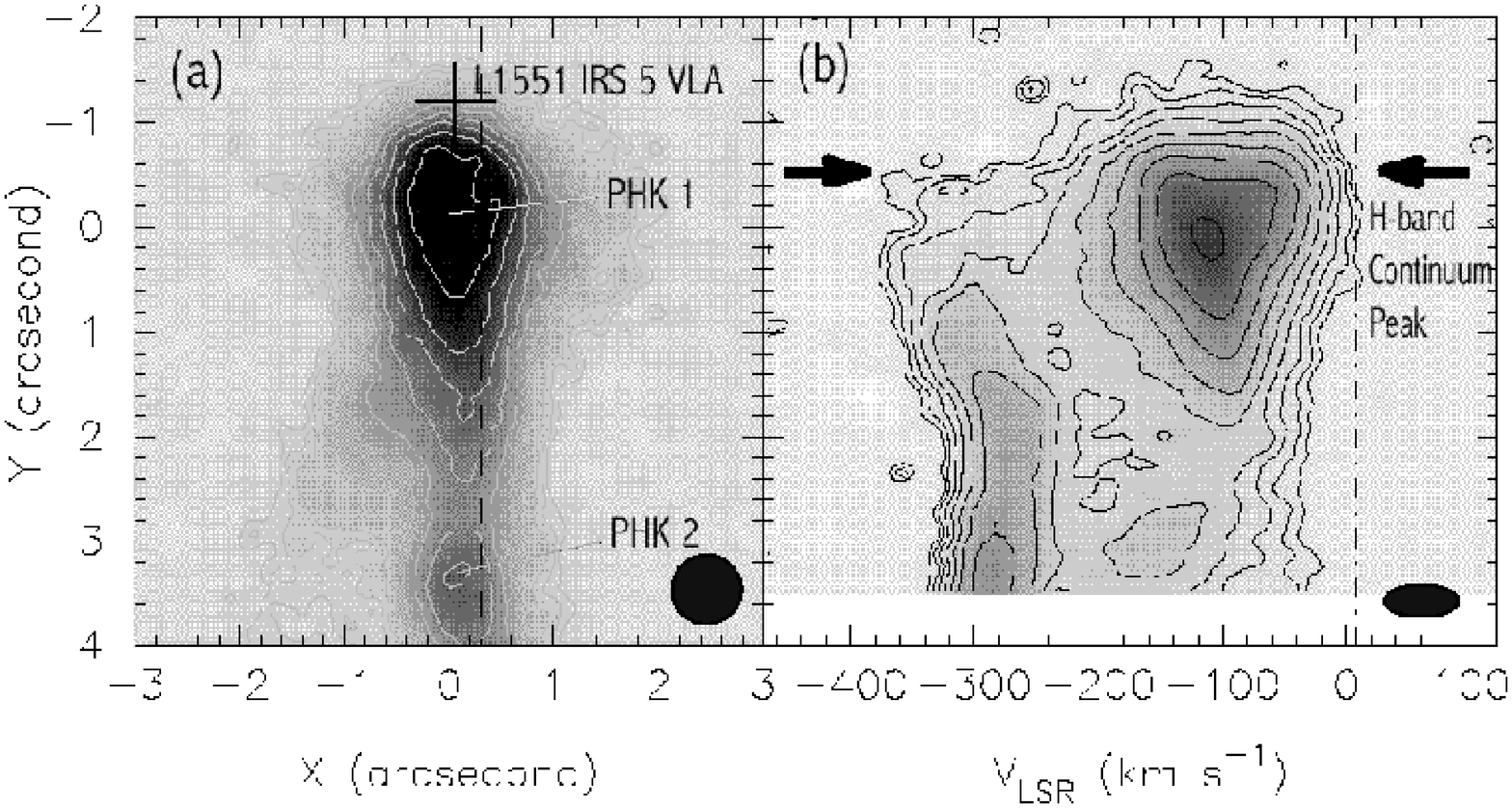}}
  \centerline{\includegraphics[height=0.99\columnwidth,angle=-90]{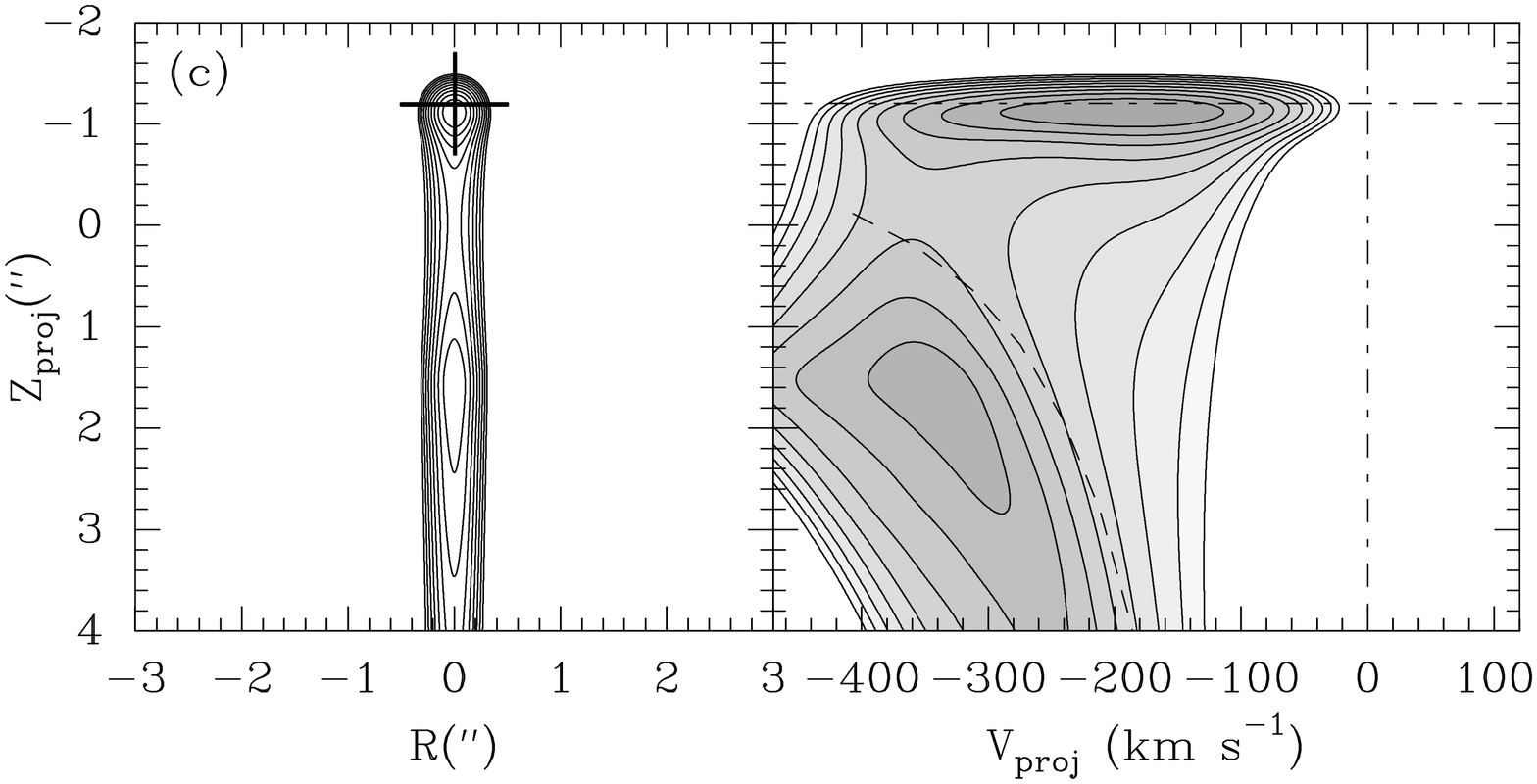}}
  \centerline{\includegraphics[height=0.99\columnwidth,angle=-90]{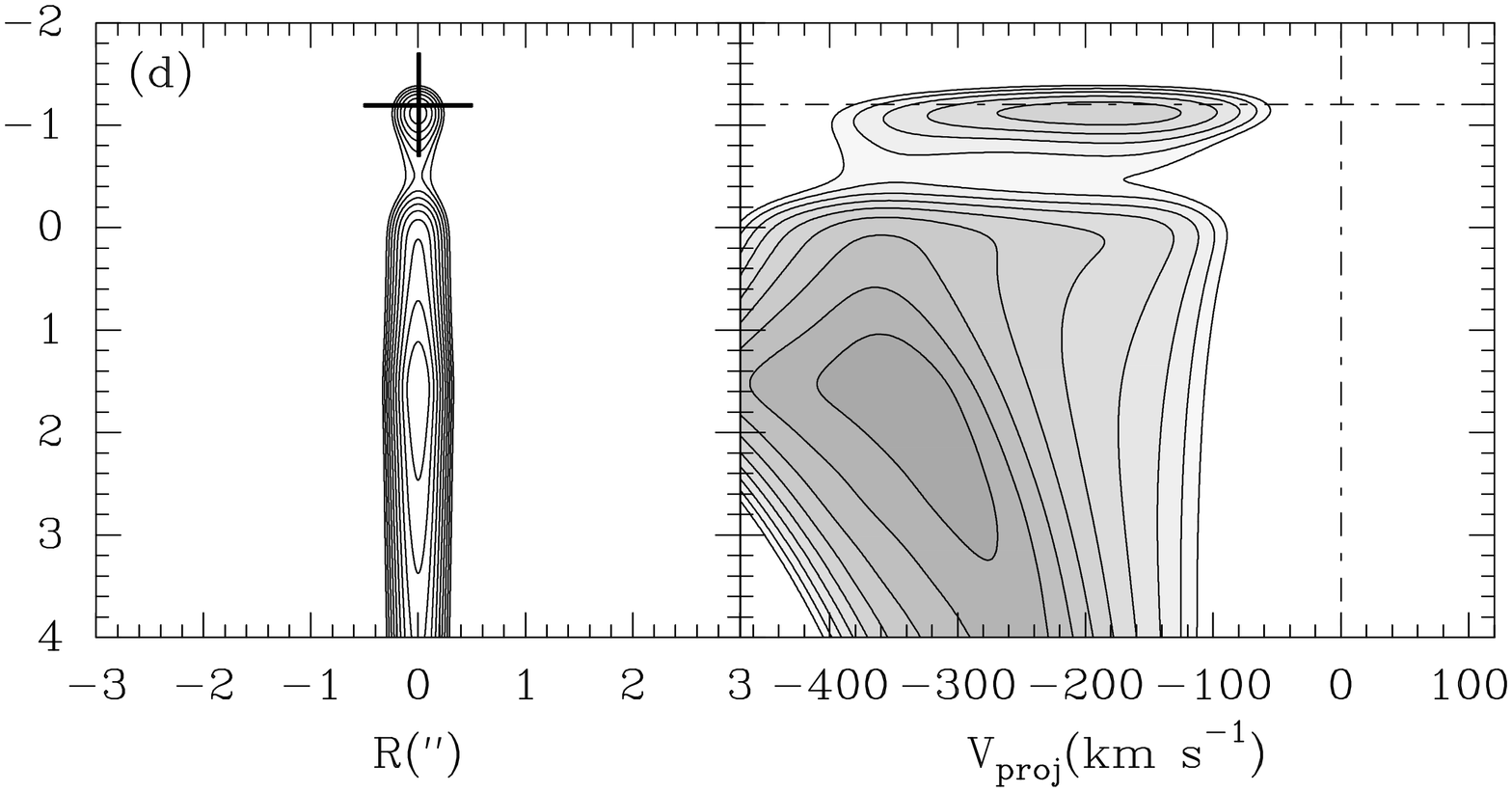}}
  \centerline{\includegraphics[height=0.99\columnwidth,angle=-90]{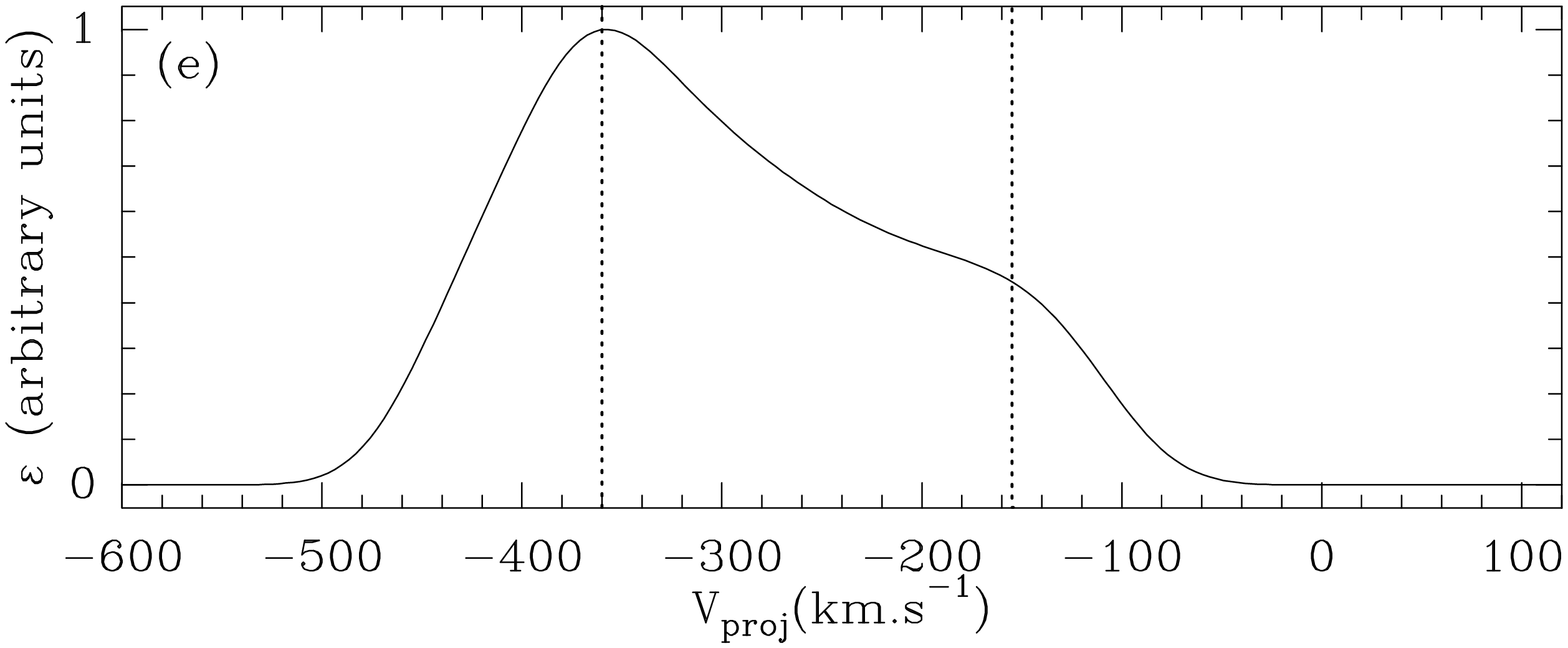}}
  \caption{\feii\ 1.644 $\mu$m emission maps ({\sl Left}) and 
    position-velocity diagrams along the jet ({\sl
    Right}). \textbf{(a),(b)}: Observations of the L1551 jet from (Pyo
    et al. \cite{pyo2002}) with a seeing size of $0.3^{\prime\prime}$
    and a velocity resolution of 59~km\,s$^{-1}$. \textbf{(c)}:
    synthetic maps predicted for a cold disk wind model with
    $\xi=0.01$, $\dot{M}_{acc} = 10^{-5}$~M$_{\odot}$\,yr$^{-1}$,
    $i=45^{\circ}$, convolved with the beam sizes of the L1551
    observations. The cross and dot-dashed lines locate the position
    of IRS5 VLA as in the top panel. The dashed line traces
    the locus where streamlines start to recollimate (beyond which our
    predictions are only approximative; see text for more
    details). \textbf{(d)}: Synthetic maps including the
    effect of an extinction screen towards L1551-IRS5. \textbf{(e)}:
    Line profile extracted from the predicted PV diagram (Panel (c))
    at the position of PHK\,1.}
  \label{pyo}
\end{figure}

Fig. \ref{pyo}a,b,c present a comparison of observed and predicted
maps in the \feii\ $1.644\ \mu$m line. We also show in Fig. \ref{pyo}d
synthetic maps taking into account the strong extinction observed
towards L1551-IRS5. We artificially dimmed the predicted maps, before
convolution with the instrumental beam, by a uniform screen of $A_v =
20$~mag extending out to $1^{\prime\prime}$, following the extinction
measurements of Itoh et al. (\cite{itoh}).

The synthetic images show that our cold disk wind model reproduces
quite well the general observed morphology (although a higher
extinction towards the source would be needed to hide completely the
base of the jet). Interestingly, the sharp decrease in extinction at
1$^{\prime\prime}$ produces an emission peak close to the observed
peak PHK\,1. We also note that refocussing onto the jet axis in our
model produces a peak at a distance comparable to the distance of knot
PHK\,2. In synthetic PV diagrams, the refocusing of streamlines
produces a strong emission ridge beyond 2$^{\prime\prime}$ from the
star, as the electronic temperature and density are locally enhanced
by compression. The apparent deceleration in this ridge is due to the
self-similarity of the solution: Streamlines of lower speeds are
launched from larger disk radii, and thus recollimate farther from the
star. A dashed curve in Fig. \ref{pyo}c shows the locus of the
points where streamlines start to recollimate. Predictions
downstream from this curve are only indicative, since gas
pressure effects (neglected in the MHD solution) should eventually
counteract compression, and start to have an important effect on the
dynamics.

Fig. \ref{pyo}e plots the synthetic line profile in the region before
jet recollimation occurs. It shows that our cold disk wind model
reproduces quite remarkably the main kinematical properties of the
central regions of the L1551 jet: the observed velocity full width at
zero intensity of 300~km\,s$^{-1}$ (350~km\,s$^{-1}$ in the model) as
well as the existence of two velocity components: a high velocity
component (HVC) and an intermediate-velocity component (IVC
\footnote{We use this denomination to differentiate the IVC from the
true low velocity component observed in classical T~Tauri stars which
have much lower centroid velocities ($-10$~km\,s$^{-1}$) and are spatially
unresolved.}) with centroid velocities $-300$~km\,s$^{-1}$
($-360$~km\,s$^{-1}$ in the model), and $-100$~km\,s$^{-1}$
($-160$~km\,s$^{-1}$ in the model) respectively. We stress in
particular that a single cold disk wind model can fully account for
the global kinematical properties, in particular the velocity extent
of the profile, due to the range of streamlines involved. Streamlines
originating from the disk inner truncation radius rapidly collimate,
producing a fast narrow jet, while outer streamlines produce a wider
angle slower wind accounting for the low-velocity part of the
profile. As already pointed out in Cabrit et al. (\cite{cabrit1999})
and Garcia et al. (\cite{garcia2001b}), a two component wind model, as
earlier suggested by Kwan \& Tademaru (\cite{kwan}), may not be
necessary to account for both observed velocity peaks (HVC and IVC).

We notice however one important discrepancy between model and
observations, apart from the apparent deceleration of the
high-velocity emission ridge discussed earlier and due to the
refocusing of streamlines towards the axis in the model. The
relative intensity of the IVC and HVC at large distances is not well
reproduced by the cold disk wind model. In the L1551 observations, the
intermediate-velocity component dominates the emission out to
3$^{\prime\prime}$ from the star, while in the model the IVC emission
drops rapidly beyond 0.5$^{\prime\prime}$. We note the possibility
that the position-velocity diagram of Pyo et al.  (\cite{pyo2002}) is
contaminated by the southern L1551 jet (with line of sight velocities
of $\simeq -60$~km\,s$^{-1}$, Hartigan et al.  \cite{hartigan2000}),
especially at intermediate-velocity towards the PHK\,1 knot where the
two jets appear to merge in the images. However, following Pyo et
al. (\cite{pyo2002}), we estimate this possibility unlikely at
distances $\geq 1^{\prime\prime}$ where the southern jet is several
times fainter than the northern one.

We now apply the same type of comparison to DG~Tau, an ``extreme''
T~Tauri star with an infrared excess intermediate between Class I and
Class II, and a particularly bright and well-studied optical jet
(Lavalley-Fouquet et al. \cite{lavalley-fouquet}; Dougados et
al. \cite{dougados}; Bacciotti et al. \cite{bacciotti2000}).  A direct
comparison with models is possible since the jet inclination is known
from proper motions, and high-resolution velocity-resolved
\feii~$1.644\ \mu$m spectra were recently published by Pyo et
al. (\cite{pyo2003}).  In DG~Tau, ejection signatures have
been shown to be highly time variable close to the central source:
Large changes in the measured velocity distribution, covering the
range from 0 to $-400$~km\,s$^{-1}$ , are observed at distances
$<$~$0.5$--$1^{\prime\prime}$ on timescales of a few years (Solf
\cite{solf1997}). However, beyond these distances, terminal velocities
varied by less than 20 \% over 10 years: $-200$~km\,s$^{-1}$ at
d~=~$3$--$3.5^{\prime\prime}$ observed in Solf \& B\"ohm
(\cite{solf1993}), $-210$ to $-190$~km\,s$^{-1}$ at
d~=~$2.5$--$4^{\prime\prime}$ in Lavalley-Fouquet et
al. (\cite{lavalley-fouquet}), $-240$~km\,s$^{-1}$ at
d~$>1.2^{\prime\prime}$ in Pyo et al. (\cite{pyo2003}).  Internal
shocks due to variability of the ejection velocity are expected to
dissipate quickly along the jet (Raga \& Kofman \cite{raga}), leaving
the average flow mostly unperturbed at large distances. Indeed,
derived shock velocities in the knots of the DG~Tau jet do not exceed
50--70~km\,s$^{-1}$ (Lavalley-Fouquet et
al. \cite{lavalley-fouquet}). A direct comparison of the properties of
the jet on scales $\geq$~100~AU with the predictions of stationary
models appears therefore justified.

\begin{figure}
  \centerline{\includegraphics[width=88mm]{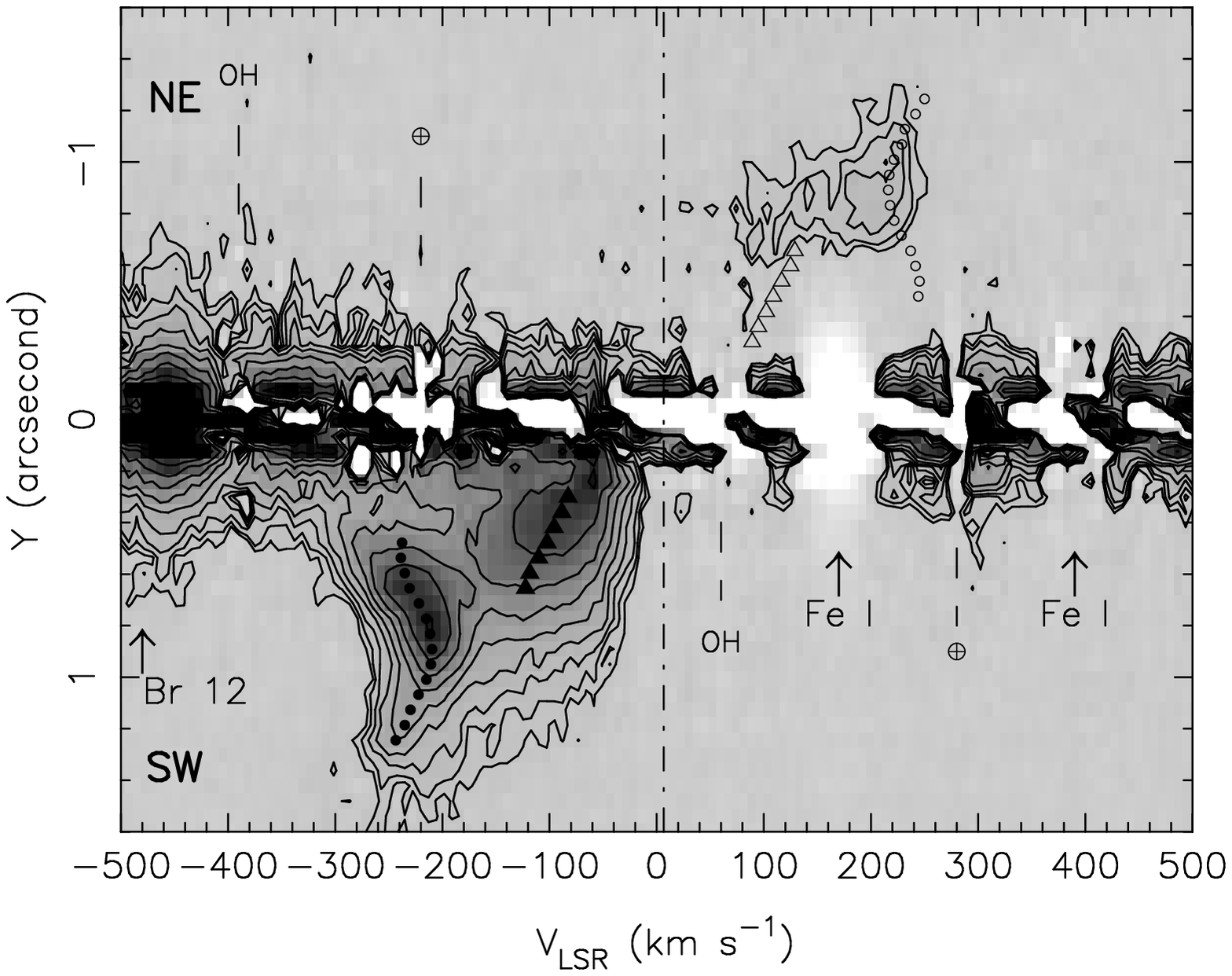}}
  \centerline{\includegraphics[height=88mm,angle=-90]{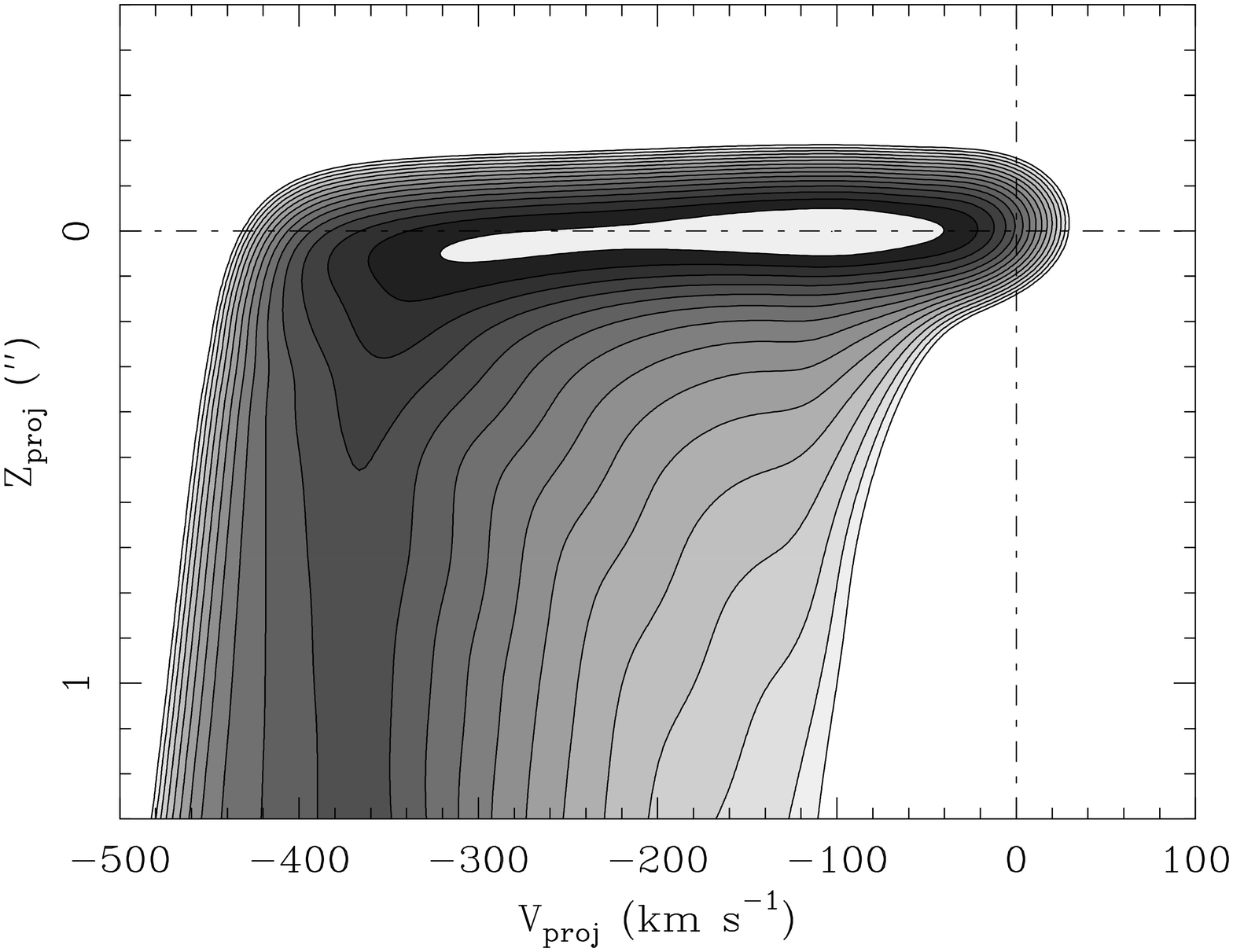}}
  \caption{ \feii~$1.644\,\mu$m position-velocity diagrams. Contours
    are drawn with equal intervals in logarithmic scale. \textbf{Top}:
    Observations of the DG~Tau jet (Pyo et al. \cite{pyo2003}) with an
    angular resolution of 0.16$^{\prime\prime}$ and a velocity
    resolution of 30~km\,s$^{-1}$. \textbf{Bottom}: Predicted diagram
    from the cold disk wind model A (Garcia et al. \cite{garcia2001b})
    with $ \xi=0.01, \dot{M}_{acc} = 10^{-6}$~M$_{\odot}$\,yr$^{-1}$,
    $i = 45^{\circ}$, convolved with the beam sizes of DG~Tau
    observations.}
  \label{DGTau}
\end{figure}

In Fig. \ref{DGTau}, we compare the long-slit spectrum from Pyo et
al. (\cite{pyo2003}) with our model predictions, convolved by the
appropriate resolution ($0.16^{\prime\prime}$ and 30~km\,s$^{-1}$). We
took an accretion rate of $10^{-6}$~M$_{\odot}$\,yr$^{-1}$, close to
the estimated value of $2 \times 10^{-6}$~M$_{\odot}$\,yr$^{-1}$
through the disk (Hartigan et al. \cite{hartigan1995}) and an
inclination angle of 45$^{\circ}$, as estimated from the proper
motions (Dougados et al. \cite{dougados}; Pyo et
al. \cite{pyo2003}). The model again explains well the presence of two
velocity components, and the general centroid velocities: the IVC
peaks at $\sim -60$ to $-130$~km\,s$^{-1}$ in DG~Tau ($-100$ to
$-140$~km\,s$^{-1}$ in the model), the HVC peaks at $-200$ to
$-240$~km\,s$^{-1}$ in DG~Tau ($-360$~km\,s$^{-1}$ in the
model). However, we note that the agreement for the HVC centroid is
less good than in L1551-IRS5, in the sense that the model now
overestimates the HVC velocity by a factor 1.5.  Another discrepancy,
also encountered in the case of L1551-IRS5, is that the IVC rapidly
becomes weaker than the HVC in our model (at
Z~$\simeq$~0.2$^{\prime\prime}$), while in the observations of Pyo et
al. (\cite{pyo2003}), the IVC dominates over the HVC much further out
from the star (until Z~$\simeq$~0.6$^{\prime\prime}$).

\subsubsection{Density}

Itoh et al. (\cite{itoh}) derive the variation of extinction and
n$_{\mathrm e}$ along the L1551 jet from JH-band spectroscopic
observations, using the \feii\ $1.257\ \mu$m/$1.644\ \mu$m and $1.644\
\mu$m/$1.533\ \mu$m line ratios (see section \ref{diag_diag}). The
derived electron densities decrease from $10^6$\ cm$^{-3}$ (at
d=2$^{\prime\prime}=280$\,AU from IRS5) to $10^4$ cm$^{-3}$
(d~=~$4^{\prime\prime}$ from IRS5). For our fiducial cold disk wind
model (with $\dot{M}_{acc}=10^{-5}$~M$_{\odot}$\,yr$^{-1}$), we
compute n$_{\mathrm e}$ from \feii\ $1.644\ \mu$m/$1.533\ \mu$m line
ratios integrated across the jet. On the same spatial scales the
predicted value is $\simeq 10^3$~cm$^{-3}$, i.e. 10 to 1000 times
smaller than observed. We note that even assuming fully ionized gas
(i.e. extra heating mechanisms: shocks or turbulent dissipation), the
density in our model would still remain too low: all streamlines have
n$_H <10^6$~cm$^{-3}$ beyond Z~$=100$~AU for
$\dot{M}_{acc}=10^{-5}$~M$_{\odot}$\,yr$^{-1}$ (see Fig. \ref{emline}
with $\dot{M}_{acc}=10^{-6}$~M$_{\odot}$\,yr$^{-1}$, and note that
n$_H\propto {\rm \dot{M}}_{acc}$). This density deficiency has already
been observed in the central regions of T Tauri microjets
(Lavalley-Fouquet et al. \cite{lavalley-fouquet}, Garcia et
al. \cite{garcia2001b}). The near-IR observations of the L1551 jet
show that self-similar {\sl cold} disk winds also fail to reproduce
the observed electronic densities in younger embedded flows on spatial
scales of a few 100~AUs.

\subsubsection{Summary}

While the near-IR \feii\ profiles predicted by the cold
disk wind model reproduce rather well the two velocity components (HVC
and IVC) observed in stellar jets, they appear to suffer from the same
deficiency of intermediate-velocity emission (as compared to
high-velocity) already noted in optical studies of the inner jet
regions in T~Tauri stars (Cabrit et al. \cite{cabrit1999}; Garcia et
al. \cite{garcia2001b}). More efficient heating at the base of
the jet appears needed to enhance the brightness of the IVC component
compared to the HVC. In addition, the predicted terminal velocity for
the HVC appears in very good agreement (within $\simeq$ 20 \%) with
the observations in the Class I source L1551-IRS5, but exceeds the
observed speed of the HVC by a factor $\simeq$~1.5 in the Class I/II
source DG~Tau. We note that the asymptotic poloidal velocity
predicted by the disk wind models can be expressed by: $v_{pol,\infty}
\simeq v_{K,0} \sqrt{2\lambda-3}$, where $v_{K,0}$ is the Keplerian
speed in the disk midplane at the cylindrical launching radius r$_0$
and $\lambda \simeq {\rm r}_{A}^2/{\rm r}_{0}^2$ the magnetic lever
arm (r$_A$ is the radius of the magnetic surface at the Alfv\'en
point). Lower terminal poloidal velocities can be achieved by
decreasing either $v_{K,0}$ (i.e. increasing r$_0$) or
$\lambda$. Larger launching radii would lead however to even lower
total jet densities, increasing the discrepancy with the
observations. Therefore, smaller magnetic lever arms (typically
$\lambda < 25$ as compared to our disk wind solution where $\lambda
\simeq 50$) are required to reproduce the observed terminal speed in
DG~Tau.

\section{Conclusions}

Thanks to a simplified 16-level atomic model, we investigated the
potential of near-IR \feii\ lines as diagnostics of physical
conditions in the inner regions of stellar jets. We presented
diagnostic diagrams that provide determinations of n$_{\mathrm e}$ in the range
$10^2$ to $10^5$~cm$^{-3}$, T$_{\mathrm e}$ below 20\,000~K, and a lower limit
to the Fe/S ratio (based on a near-infrared \feii/\sii\ line
ratio). Using observational data from Nisini et al. (\cite{nisini}) in
several HH~flows, and assuming T$_{\mathrm e} \simeq$7000~K, we deduce lower
limits to the iron gas phase abundance comparable to those found from
molecular shock models by Nisini et al. (\cite{nisini}), i.e. on
average 30\% of the solar value.  For T$_{\mathrm e} \simeq$ 10$^4$~K, the
derived Fe abundances would be roughly solar in most HH~objects,
except HH~111H and HH~241A. Hence, we confirm earlier work indicating
an efficient dust destruction mechanism in stellar jets. However, we
cannot yet conclude on the nature of the grain destruction process,
i.e. shocks versus evaporation. Calculations of grain destruction in
hydrodynamic shocks have concentrated so far on the low-density
interstellar medium (Jones et al. \cite{jones}) and similar
computations with preshock densities $\simeq 10^3-10^4$~cm$^{-3}$
typical of HH~flows are needed. Such calculations currently exist only
in the case of low-ionization magneto-hydrodynamics shocks with
ion-neutral decoupling, but this type of shocks may not occur at the
ionization level of a few \% observed in optical jets. High angular
resolution observations of the jet base in \feii\ and \sii\ lines,
e.g. with NAOS/CONICA on the VLT, are also needed to determine whether
the grains are destroyed in the vinicity of the exciting source or
further out.

In the second part of this article, we extended the work presented in
Garcia et al. (\cite{garcia2001b}) to predictions in the near-infrared
domain, by applying our Fe$^{{\rm +}}$ emissivity calculations to a
cold disk wind model with a self-consistent thermal and ionization
structure (Ferreira \cite{ferreira1997}, Garcia et
al. \cite{garcia2001a}). On large scale (a few 100~AUs), predicted
emission maps are very similar in near-IR \feii\ lines and optical
\sii\ lines. However, on scales of 1~AU soon accessible with
the AMBER/VLTI instrument, \feii\ lines appear more efficient
to trace colder and more extincted regions very close to the driving
source. In particular, AMBER/VLTI observations in \feii\ lines
are predicted to resolve the innermost streamline and inner
hole expected for a cold disk wind.

Comparing synthetic maps with recent observations of the L1551-IRS5
and DG~Tau jets in near-infrared \feii\ lines, we find that our model
is in good agreement with the global kinematics, and notably the
presence of two velocity components: a intermediate-velocity component
(IVC) at $-100$~km\,s$^{-1}$ ($-100$ to $-150$~km\,s$^{-1}$ in the
model) and a high-velocity component (HVC) at $-230$ to
$-300$~km\,s$^{-1}$ ($-360$~km\,s$^{-1}$ in the model).  However, the
predicted velocity for the HVC tends to be too high (by 20~\% only in
L1551~IRS5, but by a factor 1.5 in DG~Tau). Furthermore, the model
predicts insufficient emission at intermediate-velocity, compared to
the HVC, and also insufficient gas density in L1551. Similar failures
were encountered when comparing the cold disk wind model with
observations of jets from T Tauri stars in the optical domain (Garcia
et al. \cite{garcia2001b}).  MHD disk wind models with entropy
injection at the base of the jet (Casse \& Ferreira \cite{casse})
appear promising to solve these problems as they could provide denser
solutions, with a slower HVC and more emission from the IVC
(originating from the warm wind base). Calculations of observational
predictions for this class of ``warm disk winds'' are in progress and
will be the subject of a future paper.

\begin{acknowledgements}
We are very grateful to Anil Pradhan for communicating unpublished
results from his \mbox{142-level} code and for useful
suggestions. We would like to thank the referee, K.-H.~B\"ohm,
for his helpful comments.

\end{acknowledgements}

\end{document}